\documentclass[usenatbib]{mnras}

\usepackage{multirow}
\usepackage{rotating}
\usepackage{colortbl}
\usepackage{color}
\usepackage[fleqn]{amsmath}
\usepackage{amssymb}
\usepackage{amsfonts}
\usepackage{verbatim}
\usepackage{scalefnt}
\usepackage[percent]{overpic}
\usepackage[T1]{fontenc} 
\usepackage{aecompl} 
\usepackage{ulem}
\usepackage{cleveref}
\usepackage{pdflscape}
\usepackage{bm}
\usepackage{xfrac}
\usepackage[colorinlistoftodos]{todonotes}

\def\msun{~{\rm M}_{\sun}}
\def\lsim{\mathrel{\rlap{\lower 3pt \hbox{$\sim$}} \raise 2.0pt \hbox{$<$}}}
\def\gsim{\mathrel{\rlap{\lower 3pt \hbox{$\sim$}} \raise 2.0pt \hbox{$>$}}}

\newcommand{\comments}[1]{} 
\newcommand\T{\rule{0pt}{2.6ex}}       
\newcommand\B{\rule[-1.2ex]{0pt}{0pt}} 

\title[Feedback from spinning black holes]{Non-isotropic feedback from accreting spinning black holes}

\author[L. Sala et al.]{Luca Sala,$^{1,2}$\thanks{E-mail: l.sala40@campus.unimib.it}
Elia Cenci,$^{1}$
Pedro~R. Capelo,$^{3}$
Alessandro Lupi$^{4}$
and Massimo Dotti$^{1,5}$
\\
$^{1}$Dipartimento di Fisica ``G.~Occhialini'', Universit\`{a} degli Studi di Milano-Bicocca, Piazza della Scienza 3, I-20126 Milano, Italy\\
$^{2}$Universit\"ats-Sternwarte M\"unchen, Fakult\"at f\"ur Physik, LMU Munich, Scheinerstr. 1, D-81679 M\"unchen, Germany\\
$^{3}$Center for Theoretical Astrophysics and Cosmology, Institute for Computational Science, University of Zurich, Winterthurerstrasse\\ 190, CH-8057 Z\"urich, Switzerland\\
$^{4}$Scuola Normale Superiore, Piazza dei Cavalieri 7, I-56126 Pisa, Italy\\
$^{5}$INFN, Sezione di Milano-Bicocca, Piazza della Scienza 3, I-20126 Milano, Italy
}

\date{Accepted 2020 November 11. Received 2020 October 14; in original form 2020 July 23}

\pubyear{2020}

\begin{document}

\label{firstpage}

\pagerange{\pageref{firstpage}--\pageref{lastpage}}

\maketitle


\begin{abstract}
Active galactic nuclei (AGNs) are massive black holes (BHs) caught in the act of accreting gas at the centre of their host galaxies. Part of the accreting mass is converted to energy and released into the surrounding medium, in a process loosely referred to as AGN feedback. Most numerical simulations include AGN feedback as a sub-grid model, wherein energy or momentum (or both) is coupled to the nearby gas. In this work, we implement a new momentum-driven model in the hydrodynamics code {\textsc{gizmo}}, in which accretion from large scales is mediated by a sub-grid accretion disc model, and gas particles are stochastically kicked over a bi-conical region, to mimic observed kinetic winds. The feedback cone's axis can be set parallel either to the angular momentum of the gas surrounding the BH or to the BH spin direction, which is self-consistently evolved within the accretion-disc model. Using a circumnuclear disc (CND) as a test bed, we find that (i) the conical shape of the outflow is always visible and is weakly dependent on the launching orientation and aperture, resulting in comparable mass inflows and outflows; (ii) the cone's orientation is also similar amongst our tests, and it is not always the same as the initial value, due to the interaction with the CND playing a crucial role in shaping the outflow; and (iii) the velocity of the outflow, instead, differs and strongly depends on the interplay with the CND.
\end{abstract}



\section{Introduction}\label{sec:introduction}

Supermassive black holes (BHs) with masses in the range $10^6$--$10^{10}$~M$_{\sun}$ are typically found at the centre of massive galaxies \citep[][]{Kormendy_Richstone_1995}. They represent the most energetic and compact sources in the Universe, and are typically observed as active galactic nuclei (AGNs), through the radiation emitted by the material accreting on to them. According to the commonly accepted picture, BHs form in the early Universe and then grow with cosmic time via accretion and mergers, releasing an enormous amount of energy in the process (dubbed `AGN feedback'), that can significantly affect their surroundings. Because of their powerful feedback, AGNs are usually advocated to explain the quenching of star formation in the most massive galaxies \citep{Silk_and_Rees_1998,DiMatteo_et_al_2005}, and have been also proposed as potential candidates to solve other open problems, such as the source of hydrogen reionization \citep{Volonteri_and_Gnedin_2009}, the too-big-to-fail problem \citep{GarrisonKimmel_et_al_2013}, and the removal of gas from massive disc progenitors \citep[e.g.][]{Peirani_et_al_2012}, although a clear consensus has not been reached yet.

AGN feedback, because of its role in the BH mass growth self-regulation, is also invoked as a mechanism to explain the emergence of the observed correlations between the central BH and the velocity dispersion of the central stellar spheroid \citep[e.g.][]{Ferrarese_Merritt_2000,Gebhardt_et_al_2000,Haring_Rix_2004,Gultekin_et_al_2009,Kormendy_Ho_2013} or its stellar mass \citep[e.g.][]{Magorrian_et_al_1998,McConnell_and_Ma_2013,Reines_and_Volonteri_2015}. In particular, different scenarios have been proposed to date to explain these correlations, wherein (i) the galaxy grows first and the BH catches up later in a supply-limited fashion, (ii) the BH is initially overmassive and the stellar component is built-up only after the AGN feedback-regulated regime has been reached, or (iii) stars and BH co-evolve along the correlation in a self-regulated fashion \citep[see reviews by e.g.][]{Volonteri_2012,Kormendy_Ho_2013}.  Despite the increasing amount of data available, and the information obtained at high redshift, a clear consensus about how and when these correlations arose is still missing.

In this respect, AGN feedback can play a crucial role, and the observation of jets and large-scale winds represents an extremely important evidence of this mechanism at play. In particular, fast nuclear (sub-pc) outflows \citep{Tombesi_et_al_2013,Nardini_et_al_2015} as well as galaxy-scale (kpc) winds \citep{Feruglio_et_al_2010,Rupke_2011,Sturm_et_al_2011, Greene_et_al_2012,Maiolino_et_al_2012,Liu_et_al_2013,Cicone_et_al_2014,Harrison_et_al_2014} allow one to infer how much energy and momentum are injected in the BH surroundings and transported up to kpc scales, affecting the evolution of the system, and can be considered as the indirect evidence of the existence of such feedback processes.

In the X-ray band \citep{Gofford_2015}, ultrafast outflows (with velocities larger than 10 per cent of the speed of light) are observed in 40 per cent of the bright local population of AGNs. Deep absorption lines associated with highly ionized iron \citep{Reeves_et_al_2003,Tombesi_et_al_2013,Gofford_2015} are also typically found to be blueshifted with velocities in the range $\sim$10$^3$--10$^4$~km~s$^{-1}$, suggesting the presence of outflowing material directly connected to the central BH energy injection on sub-pc scales. From the detection fraction of 0.4, an opening angle can also be derived for these fast pc-scales winds \citep[as done in e.g.][]{Gofford_2015}, corresponding to a region encompassed by the outflow which subtends almost half of a sphere.

Recently, massive galaxy-size cold molecular outflows (up to $10^{10}\msun$ in molecular hydrogen) have been detected with velocities in the range $\sim$10$^2$--10$^3$~km~s$^{-1}$ at far-infrared and millimetre wavelengths \citep[e.g.][]{Feruglio_et_al_2010,Fischer_2010,Cicone_et_al_2014,Cicone_et_al_2018,Garcia-Burillo_2014,Sirressi_et_al_2019}. In addition, there are examples of winds associated with molecular gas, with speeds of the order of 10$^3$~km~s$^{-1}$, showing both collimated and un-collimated geometries at hundreds of pc scales \cite[see e.g.][and references therein]{Combes_2017}.

Theoretically, several works have addressed the role of BH feedback in the emergence of galaxy--BH correlations and in galaxy quenching  \citep[e.g.][]{Silk_and_Rees_1998,Fabian_1999,King_2003}, showing that, when the luminosity produced by the central object couples with the nearby matter, a fast nuclear wind is generated, which can interact with the interstellar medium (ISM) of the host, sweeping up material and driving large-scale outflows able to suppress star formation. In addition, by accounting for the energetics of the shock-generated shell after an AGN feedback event, the scaling relation between the stellar spheroid's velocity dispersion $\sigma$ and BH mass $M_{\rm BH}$ can be derived \citep{King_2003} either in the fully energy-conserving regime ($M_{\rm BH}\propto \sigma^5$) or in the momentum-conserving case ($M_{\rm BH}\propto \sigma^4$), bracketing extremely well the observed correlation, i.e. $M_{\rm BH}\propto \sigma^{4.38}$ \citep{Kormendy_Ho_2013}.

However, given the complexity of the AGN feedback problem and its interplay with the galaxy host, involving a non-linear coupling of multiple physical processes on extremely different scales (from the accretion disc up to the cosmological environment), a careful combination of observations, analytical modelling, and numerical simulations is needed. Over the last decade, many numerical hydrodynamic simulations have tackled this problem, focussing on different scales. For instance, at small scales, several studies have investigated the effects of radiation on the accretion itself and the wind/jet driving with radiation-hydrodynamics -- in some cases, also including magnetic fields and general relativistic effects \citep[e.g.][]{Park_and_Ricotti_2012,Sadowski_2016,Jiang_et_al_2019}. At larger scales (and lower resolution), where the accretion disc scales cannot be properly resolved, effective sub-grid models were developed, mostly to address the effects of AGN feedback on the galaxy ISM and the BH dynamics \citep[see  e.g.][]{Hopkins_et_al_2007,Choi_et_al_2012,Capelo_et_al_2015,Zubovas_et_al_2016,Angles-Alcazar_et_al_2017,SouzaLima_et_al_2017, Lupi_et_al_2019}. Finally, at cosmological scales, where the spatial resolution is too low to properly resolve the galaxy ISM, AGN feedback is mainly modelled via an effective coupling efficiency of the accretion-powered luminosity with the gas, calibrated to reproduce the observed local correlations, as in e.g. \citet{Schaye_et_al_2015,Dubois_et_al_2014} and \cite{Vogelsberger_et_al_2014}.

Although most of the current state-of-the-art cosmological simulations have been shown to correctly reproduce many galaxy and quasar properties, we are still far from a proper understanding of how the AGN feedback actually couples on different scales and at different resolutions. For example, alternative explanatory scenarios for the correlations we observe are possible: \citet*{Angles_Alcazar_et_al_2013} and \cite{Angles_Alcazar_et_al_2015}, using the sub-resolution accretion rate estimator based on gravitational torques by \cite{Hopkins_Quataert_2011}, showed that feedback-regulated growth is not required for the emergence of scaling relations between the BH and its host galaxy.

In this work, we build upon the BH accretion and feedback models by \citet{Angles-Alcazar_et_al_2017} and introduce a self-consistent model of a sub-resolution \citet{Shakura_and_Sunyaev_1973} accretion disc that evolves via accretion from the surroundings (on resolved scales), draining due to the gas accretion on to the central BH, and wind production (i.e. outflows). We implement our model in the publicly available code {\textsc{gizmo}}\footnote{\url{http://www.tapir.caltech.edu/~phopkins/Site/GIZMO.html}} \citep{Hopkins_2015}, descendant of {\textsc{gadget2}} and {\textsc{gadget3}} \citep{Springel_2005,Springel_2008}, and validate it via a suite of circumnuclear disc (CND) simulations. We also explore the role of the central BH spin, by coupling our sub-grid model with the self-consistent spin evolution model presented in \cite{Cenci_et_al_2021}.

The paper is organized as follows: in Section~\ref{sec:implementation}, we describe our accretion disc model and the coupling with the hydrodynamic code; in Section~\ref{sec:tests}, we validate the model in a controlled environment; and in Section~\ref{sec:conclusions}, we draw our conclusions.


\section{Implementation}\label{sec:implementation}

In this section, we describe the sub-grid model for BH accretion and feedback employed in this work, which builds upon the one presented in \citet{Angles-Alcazar_et_al_2017}, where we add the sub-resolution modelling of an accretion disc and allow the outflow to have different shapes and directions. In the following, we provide the general description of the model, whereas all specific values of the parameters are given in Section~\ref{sec:tests}.


\subsection{Accretion from large scales}\label{sec:accretion_from_large_scales}

Since galaxy- and cosmological-scale simulations are unable to properly resolve the accretion disc around BHs (of the order of the Schwarzschild radius), accretion is usually modelled via semi-analytical prescriptions that link the resolved inflow rate to the actual accretion on to the BH.

In \textsc{gizmo}, the mass transfer rate from resolved scales onto a BH particle (corresponding to a sub-resolution `BH+accretion disc' system) can be estimated using a variety of different prescriptions.
In this work, we assume that the mass transfer rate on to the unresolved system is determined by the Bondi--Hoyle--Lyttleton formula \citep[hereafter Bondi;][]{Hoyle_Lyttleton_1939,Bondi_Hoyle_1944,Bondi_1952} as implemented by \citet{Springel_et_al_2005}:\footnote{We note that, a priori, any other prescription to estimate the inflow rate would be equally valid. Furthermore, being the Bondi accretion employed to estimate the inflow on to the BH+disc system, $M_{\rm BH}$ should be the sum of the BH and accretion disc masses. However, we opted for simplicity to use the already existing code module without any modification, motivated by the fact that, in practice, the mass of the accretion disc is $\lesssim 10^{-2}\; M_{\rm BH}$.}

\begin{equation}
    \dot{M}_{\rm in}=\frac{4\pi \alpha_{\rm acc} G^2 M^2_{\rm BH} \rho}{(c_{\rm s}^2+v^2)^{3/2}}, \label{eq:bondi}
\end{equation}

\noindent where $\rho$ and $c_{\rm s}$ are the density and the sound speed of the gas, respectively, evaluated as mass-weighted average quantities using the BH neighbouring gas particles, $v$ is the velocity of the BH relative to the centre-of-mass velocity of the gas within the BH kernel length, $M_{\rm BH} = 10^6 M_{\rm BH,6}$~M$_{\sun}$ is the BH mass, $G$ is the gravitational constant, and $\alpha_{\rm acc}$ is a dimensionless parameter usually employed in $\sim$kpc-resolution simulations to boost the accretion rate in the aim at correcting for the unresolved dense gas in the ISM \citep{DiMatteo_et_al_2005,Booth_and_Schaye_2009}. Due to the resolution of our tests, $\alpha_{\rm acc}$ is set to 1.

The local properties of the gas surrounding the BH are averaged via a smoothing kernel defined as the region encompassing an effective number of neighbours $N_{\rm ngb,BH}=f_{\rm BH}N_{\rm ngb}$, where $N_{\rm ngb}$ is the user-defined number of neighbours used for hydrodynamics and $f_{\rm BH}\geq 1$ is a free parameter (usually chosen to ensure a good covering of the region surrounding the BH). In order to prevent the BH from un-physically accreting gas at large distances, the code allows to define a maximum kernel size used for accretion, i.e. the maximum accretion radius $r_{\rm accr,max}$, by means of an additional free parameter.

\subsection{The original model}\label{sec:the_original_model}

In the model by \citet{Angles-Alcazar_et_al_2017}, two distinct masses are associated with the BH particles, dubbed the `physical' and the `dynamical' mass, respectively. The former is evolved by any chosen sub-grid accretion model in order to follow the BH mass growth, whereas the latter is the mass that is used to calculate the BH's gravitational force and dynamics.

The BH physical mass, $M_{\rm BH}$, evolves at each time-step as

\begin{equation}\label{eq:mphysupdate}
\begin{split}
M_{\rm BH}'\equiv &M_{\rm BH}(t+\Delta t)=\\=&M_{\rm BH}(t)+\Delta M_{\rm BH}=M_{\rm BH}(t)+\dot{M}_{\rm BH}\Delta t,
\end{split}
\end{equation}

\noindent where

\begin{equation}\label{eq:BH_accr_eta}
\dot{M}_{\rm BH}=\left(1-\eta\right)\dot{M}_{\rm acc-BH},
\end{equation}

\noindent $\dot{M}_{\rm acc-BH}$ is the accretion rate on to the BH (see below), $\Delta t$ is the BH time-step in the simulation, and $\eta = 0.1\eta_{0.1}$ is the radiative efficiency, which can be kept constant during the simulation by setting its value in the parameters file of the code, or let free to vary self-consistently with the properties of the sub-grid accretion disc, as done in, e.g. \cite{Cenci_et_al_2021}.

In the original model, the accretion on to the BH is not mediated by an accretion disc. Therefore, the accretion rate, $\dot{M}_{\rm acc-BH}$, is assumed to be a fixed fraction $f$ of the large-scale inflow rate:

\begin{equation}\label{eq:MdotaccBH_alcazar}
    \dot{M}_{\rm acc-BH}=f\dot{M}_{\rm in}.
\end{equation}

\noindent Mass conservation imposes 

\begin{equation}
    \dot{M}_{\rm in}=\dot{M}_{\rm acc-BH}+\dot{M}_{\rm out},
\end{equation}

\noindent where $\dot{M}_{\rm out}$ is the outflow rate. Hence

\begin{align}\label{eq:mass_loading}
    \frac{1-f}{f}=\frac{\dot{M}_{\rm out}}{\dot{M}_{\rm acc-BH}}.
\end{align}

In a radiatively driven-wind scenario, gas is accelerated by radiation pressure, with the photons being produced during the accretion process yielding a total (bolometric) luminosity $L_{\rm bol}=\eta \dot M_{\rm acc-BH} c^2$, where $c$ is the speed of light in vacuum. The outflow rate is then given by the definition of outflow momentum flux in units of $L_{\rm bol}/c$ (hereafter ``momentum-loading''):

\begin{equation}
    p_{\rm b}=\frac{\dot{M}_{\rm out}v_{\rm out}}{L_{\rm bol}/c}=\frac{v_{\rm out}}{\eta c}\frac{\dot{M}_{\rm out}}{\dot{M}_{\rm acc-BH}},
    \label{eq:mom_load}
\end{equation}

\noindent where $v_{\rm out}$ is the outflow velocity's magnitude. The quantities $v_{\rm out}$ and $p_{\rm b}$ are free parameters of the model and can be set in the parameters file of the code.

Equations~\eqref{eq:mass_loading} and \eqref{eq:mom_load} provide a definition for $f$:

\begin{align}\label{eq:f}
    f=\frac{1}{1+(p_{\rm b}\eta c/v_{\rm out})}.
\end{align}

Once the physical BH mass growth has been computed, a set of gas particles are stochastically selected within the BH kernel, assigning to each neighbour a probability

\begin{equation}
\begin{cases}
    p_{j}=\max\Bigl\{\Bigl[(M_{\rm BH}'-M_{\rm BH_{dyn}})\frac{1}{f}&-\sum_k{m_{k}}\Bigr] \frac{w_{j}}{\rho},0\Bigr\}\\& \text{if } M_{\rm BH}'\geq M_{\rm BH_{dyn}},\\
    p_{j}=\frac{1-f}{f} \;\; \frac{w_{j}}{m_{j}} \;\; \dot{M}_{\mathrm{acc-BH}} \Delta t&\\ & \text{if } M_{\rm BH}'<M_{\rm BH_{dyn}},
\end{cases}
\label{eq:simplepj}
\end{equation}

\noindent where $M_{\rm BH_{\rm dyn}}$ is the BH dynamical mass at the beginning of the time-step, $w_{j}$ is the kernel weight, $m_{k}$ is the gas particle mass, and the sum is over the particles that have already been selected in the same time-step.

In the case $M_{\rm BH}'>M_{\rm BH_{dyn}}$, the factor $1/f$ accounts for the mass that has to be ejected. After being selected, a gas particle has a fraction $f$ of the mass removed and transferred to the BH dynamical mass,

\begin{equation}\label{eq:mdynupdate}
M_{\rm BH_{dyn}}(t+\Delta t)=M_{\rm BH_{dyn}}(t)+\sum_j f m_{j},
\end{equation}

\noindent such that, stochastically,

\begin{equation}\label{eq:simpledynmass}
\Delta M_{\rm BH_{dyn}} \simeq \Delta M_{\rm BH}.
\end{equation}

Hence, the BH dynamical mass keeps track of the discrete accretion, i.e. it increases by an amount equal to each accreted particle in the simulation (or a fraction of it). The physical mass, on the other hand, keeps track of the quasi-continuous accretion (that evolves with the BH time-step of the simulation, $\Delta t$).

In the case $M_{\rm BH}'<M_{\rm BH_{dyn}}$, we only change the momentum of the selected particles, and the BH's dynamical mass does not change. This ensures mass conservation and, at the same time, that the dynamical mass follows the physical mass (on average).

In both cases, a velocity with magnitude $v_{\rm out}$ is added to the velocity of each particle that has been selected in a specific direction, hence giving it a ``kick'' in that direction. Concerning the specific prescription for the direction, the implementation by \citet{Angles-Alcazar_et_al_2017} assumes that the kicks can be given either along the radial direction connecting the gas particle to the BH (`radial outflow') or along the angular momentum vector of the gas particles within the BH kernel length  (`collimated outflow').

We note that, in this model, the overall amount of ejected mass per time-step is stochastically equal to $\dot{M}_{\rm out}\Delta t$, and $\dot{M}_{\rm out}$ is defined once $f$ -- which depends only on free parameters and remains constant -- has been provided.


\subsection{Self-consistent inclusion of the accretion disc}\label{sec:discaccr}


\subsubsection{The sub-grid model}\label{sec:sub-grid-disc}

In our implementation, we add a self-consistent sub-resolution model of an accretion disc. Since mass must be conserved for both resolved and unresolved structures, we write the rate of change of the disc mass, $M_{\rm disc} = 10^4 M_{\rm disc,4}$~M$_{\sun}$, as

\begin{equation}
    \dot M_{\rm disc} = \dot M_{\rm in}-\dot M_{\rm acc-BH}-\dot M_{\rm out}, \label{eq:mdotdisc}
\end{equation}

\noindent where the inflow rate is computed from resolved large-scale quantities according to Equation~\eqref{eq:bondi}, and the outflow rate is computed as in Equation~\eqref{eq:mom_load}.

The accretion rate on to a BH is computed as $\dot M_{\rm acc-BH} = f_{\rm Edd} \dot{M}_{\rm Edd}$, where $\dot{M}_{\rm Edd} = 4\pi G M_{\rm BH} m_{\rm p}/(\sigma_{\rm T}\eta c)$ is the \citet{Eddington_1916} accretion rate, $m_{\rm p}$ the proton mass, and $\sigma_{\rm T}$ the Thomson cross section. The Eddington ratio $f_{\rm Edd}$ is given by \citep[][]{Fiacconi_et_al_2018}

\begin{equation}\label{eq:fedd}
    f_{\rm Edd} \simeq \; 0.76 \; \eta_{0.1} \; \alpha_{0.1}^{8/7}\; M_{\rm disc,4}^{5} \;M_{\rm BH,6}^{-47/7} \left( \frac{a_{\rm BH}}{3}\frac{J_{\rm disc}}{J_{\rm BH}} \right)^{-25/7},
\end{equation}

\noindent where $a_{\rm BH} = c J_{\rm BH} /(G M_{\rm BH}^2)$ is the BH spin parameter, $J_{\rm disc}$ is the angular momentum magnitude of the accretion disc, $\alpha = 0.1 \alpha_{0.1}$ is the disc viscosity parameter, and $J_{\rm BH}$ is the magnitude of the BH angular momentum, defined as $\bm{J}_{\rm BH}= J_{\rm BH}\cdot\bm{j}_{\rm BH}$, where $\bm{j}_{\rm BH}$ is its direction. Furthermore, note that, in our model $\eta$ is computed as a function of $a_{\rm BH}$. We further impose that $f_{\rm Edd} \leq 1$ and, at every time-step, we limit $\dot{M}_{\rm in}$ to prevent the sub-grid disc from becoming self-gravitating (for more details on how $\alpha$, $\eta$, and the limitation on $\dot{M}_{\rm in}$ are computed, see \citealt{Cenci_et_al_2021}). Finally, at any given time-step, if the mass subtracted from the disc is greater than the disc mass itself, the disc mass is set to zero.\footnote{GIZMO already implements a time-step criterion for BH accretion, based on the accretion rate computed during the last BH step, aimed at ensuring that the BH does not accrete more the 0.1 per cent of its mass. Although this criterion is already sufficient to almost exclude the disc depletion in most of the cases (also because of the decay of the accretion rate when the disc mass approaches zero), some exceptions could still occur that cannot be captured by any criteria based on the previous time-step values, and these occur because of the complex dependence of the disc mass evolution on the inflow and outflow rates at each time-step.}


\subsubsection{Stochastic accretion and feedback}\label{sec:stochastic_accretion_and_feedback}

We evolve the BH physical mass as in Equations \eqref{eq:mphysupdate} and \eqref{eq:BH_accr_eta}. We additionally follow the accretion disc mass that is updated at each time-step as

\begin{align}\label{eq:mdiscupdate}
\begin{split}
M_{\rm disc}'\equiv &M_{\rm disc}(t+\Delta t)=\\=&M_{\rm disc}(t)+\Delta M_{\rm disc}=M_{\rm disc}(t)+\dot{M}_{\rm disc}\Delta t=\\=&M_{\rm disc}(t)+\dot{M}_{\rm in}\Delta t-\dot{M}_{\rm acc-BH}\Delta t-\dot{M}_{\rm out}\Delta t,
\end{split}\\ \nonumber
\end{align}

\noindent where $\dot{M}_{\rm disc}$ is given by Equation~\eqref{eq:mdotdisc}.

The stochastic selection probability in Equation~\eqref{eq:simplepj} must therefore be modified as

\begin{equation}
\begin{cases}
    p_{j}=\max\Biggl\{0, \Biggl[M_{\rm BH}'+M_{\rm disc}'\\+\dot{M}_{\rm out}\Delta t\\-\Biggl(M_{\rm BH_{dyn}}+\sum_k{m_{k}}\Biggr)\Biggr] \frac{w_{j}}{\rho}\Biggr\}&\text{if } M_{\rm BH}'+M_{\rm disc}'\geq M_{\rm BH_{dyn}},\\
    p_{j}=\frac{w_{j}}{m_{j}} \;\; \dot{M}_{\rm out} \;\; \Delta t&\text{if } M_{\rm BH}'+M_{\rm disc}'<M_{\rm BH_{dyn}}.
\end{cases}
\label{eq:newpj}
\end{equation}

In the case $M_{\rm BH}'+M_{\rm disc}'>M_{\rm BH_{dyn}}$, $\dot{M}_{\rm out}\Delta t$ accounts for the mass left to the particles after accretion, that will constitute the outflow mass. The mass fraction that has to be removed from each particle is consistently re-defined as

\begin{equation}\label{eq:newfraction}
    f=1-\frac{\dot{M}_{\rm out}\Delta t}{\sum_{k=1}^N m_k},
\end{equation}

\noindent where $N$ is the total number of selected particles. The code removes this mass fraction from each particle and adds it to $M_{\rm BH_{dyn}}$, as in Equation~\eqref{eq:mdynupdate}, such that, stochastically,

\begin{equation}\label{eq:dynmass}
\Delta M_{\rm BH_{dyn}} \simeq \Delta M_{\rm BH} + \Delta M_{\rm disc} \simeq \dot{M}_{\rm in}\Delta t-\dot{M}_{\rm out}\Delta t.
\end{equation}

Next, each particle is kicked with its remaining mass, which is then just $\Delta M_{\rm out}/N$. We stress that our redefinition of the accreted fraction from each particle is variable on a time-step basis, because it depends on the number of selected particles, which in turn depends on both the amount of mass that has to be accreted and the amount launched in the outflow. This redefinition is necessary because of the presence of the sub-grid accretion disc. When an accretion disc is present, the constant $f$ in the previous model is not valid any longer, since $\dot{M}_{\rm out}$ is defined by the accretion rate on to the BH, and not by that on to the disc. Hence, $f$ is defined only once $\dot{M}_{\rm out}$ has been computed starting from $\dot{M}_{\rm acc-BH}$. However, the particles are selected such that the whole amount of ejected mass per time-step is stochastically equal to  $\dot{M}_{\rm out}\Delta t$.

In the case $M_{\rm BH}'+M_{\rm disc}'<M_{\rm BH_{dyn}}$, $\Delta M_{\rm BH_{dyn}} = 0$ and we only change the momentum of the selected particles.

\subsection{Conical feedback}\label{sec:conical_feedback}

In order to better reflect observations of radiatively driven winds \citep[see  e.g.][]{Gofford_2015,Combes_2017}, we implement an outflow with a biconical shape, whose axis can be either fixed in time (to a given arbitrary direction) or evolved during the simulation (e.g. following the angular momentum of the gas surrounding the BH). We further model the case in which the cone's axis is set parallel to the evolving BH spin, given by the prescription described in \cite{Cenci_et_al_2021}.

For each particle within the BH kernel that receives a kick, we randomly sample the kick direction within a cone of semi-aperture $\hat\theta$ aligned with a chosen direction (either fixed or time-dependent). Hereafter, we will denote this direction, along which feedback is exerted, using a spherical coordinates system ($\theta$, $\phi$), respectively the polar and azimuthal angle. Note that, in our model, all the particles within the BH kernel are eligible to be kicked: only the direction of the kick is extracted such that the angles distribution describe a cone in the velocities space. This translates into a conical outflow assuming the shape of a cone (in the positions space) with the base corresponding to the BH kernel. This approach has the advantage of always finding particles to kick, unlike other schemes where, e.g. only particles within the cone are selected, possibly resulting in missing feedback \citep[][]{Barai_et_al_2016}.

In order to produce a symmetric outflow with respect to the plane perpendicular to the chosen axis, the algorithm then proceeds in two steps. First, it computes the radial position of each particle with respect to the BH, $\bm{r}$, and the scalar product of this vector and the randomly extracted velocity vector, $\bm{r}\cdot\bm{v}$. If the latter quantity is positive, then the direction is left unchanged, otherwise $\bm{v}\rightarrow -\bm{v}$. With this strategy, the volume is divided in two semispaces: the semispace where the angle between the position vector with the BH as origin and the sampled direction is less than $\pi/2$, and the complementary of this space. The particles belonging to the former semispace are kicked ``upwards'', the others are kicked ``downwards''. Statistically, the particles are split in two equally populated groups and the outflow is symmetric as a result.

\begin{table}
\centering
\caption{Resolution of our simulation suite. For each particle type, $N$ is the number of particles, $m$ (or $M$) the particle mass (in M$_{\sun}$), and $\varepsilon$ the gravitational softening (in pc). While the stellar quantities are fixed during the simulations, the gas and BH particle masses (as well as the number of gas particles) evolve, as a result of accretion events, hence we report here the initial values (with the subscript 0; also, $M_{\rm BH,0} + M_{\rm disc,0} = M_{\rm BH_{\rm dyn},0}$, or $M_{\rm BH,0} = M_{\rm BH_{\rm dyn},0}$ when the accretion disc model is not used). The softening is fixed for stars and the BH, whereas fully adaptive softening is employed for gas, whose minimum is here reported.}
\begin{tabular}{c|cc}
\hline
Resolution:                                                                & Low                           & High                             \T \B \\ \hline
$N_{\rm stars}$, $N_{\rm gas,0}$, $N_{\rm BH}$                               & $10^5$, $10^5$, 1              & 8$\times$$10^5$, 8$\times$$10^5$, 1 \T \B \\
$m_{\rm stars}$, $m_{\rm gas,0}$, $M_{\rm BH,0}$                               & 5$\times$$10^3$, $10^3$, $10^7$ & 625, 125, $10^7$                 \T \B \\
$\varepsilon_{\rm stars}$, $\varepsilon_{\rm gas}$, $\varepsilon_{\rm BH}$ & 0.16, 0.16, 1                 & 0.08, 0.08, 1                    \T \B \\
\end{tabular}
\label{tab:resolution}
\end{table}


\section{Tests}\label{sec:tests}

\begin{table*}
\centering
\caption{Summary of the parameters adopted in our simulations. The subscript 0 refers to quantities evaluated at initialization, i.e. at time t = 0. Moreover, the following parameters are the same for all tests: $p_{\rm b}=1$, $\eta_{0}=0.1$ and, for the accretion disc model in the last four runs, $a_{\rm BH,0}=0.5$, $M_{\rm disc,0}=5\times 10^4 \msun$,  $f_{\rm Edd, 0} = 5\times10^{-3}$, and $R_{\rm circ}/R_{\rm sg} = 0.5$.}
\begin{tabular}{c|cccccc}
Simulation name            & $\hat\theta$       & $\theta_{0}$ & $v_{\rm out}$ (km~s$^{-1}$) & $f$      & $\eta$   & Feedback direction     \T \B \\ \hline
\verb|radial_v1e3|         & Not applicable     & n.a.         & $10^3$                      & 0.032    & 0.1      & n.a.                   \T \B \\
\verb|collimated_v1e3|     & n.a.               & 0            & $10^3$                      & 0.032    & 0.1      & $\bm{J}_{\rm gas}$     \T \B \\
\verb|conical_v1e3|        & $\upi/4$           & 0            & $10^3$                      & 0.032    & 0.1      & $\bm{J}_{\rm gas}$     \T \B \\
\verb|radial_v1e4|         & n.a.               & n.a.         & $10^4$                      & 0.25     & 0.1      & n.a.                   \T \B \\
\verb|collimated_v1e4|     & n.a.               & 0            & $10^4$                      & 0.25     & 0.1      & $\bm{J}_{\rm gas}$     \T \B \\
\verb|conical_v1e4|        & $\upi/4$           & 0            & $10^4$                      & 0.25     & 0.1      & $\bm{J}_{\rm gas}$     \T \B \\
\verb|along_z_45|          & $\upi/4$           & 0            & $10^3$                      & Variable & Variable & $z$-axis               \T \B \\
\verb|along_jBH0_45| & $\upi/4$           & $5\upi/6$    & $10^3$                      & Variable & Variable & $\bm{j}_{\rm BH,0}$    \T \B \\
\verb|along_jBH0_10|       & \textbf{$\upi/18$} & $5\upi/6$    & $10^3$                      & Variable & Variable & $\bm{j}_{\rm BH,0}$    \T \B\\
\verb|full_model|          & $\upi/4$           & $5\upi/6$    & $10^3$                      & Variable & Variable & $\bm{j}_{\rm BH}$      \T \B
\end{tabular}
\label{tab:table2}
\end{table*}
 
In order to study the accretion and feedback processes in detail, we chose a test setup that allowed us to have a well-controlled environment, with the minimum amount of active sub-grid physics. The tests were performed employing the initial condition prescriptions from \cite{Lupi_et_al_2015}. The setup has three components:

\begin{itemize}

    \item a stellar spherical structure described by a \citet{Hernquist_1990} profile with total mass $M_{\rm b}=5\times10^8\msun$ and scale radius $r_{\rm b}=100$~pc, defined by the radial density profile
    
    \begin{equation}
    \rho_{\rm b}(r)=\frac{M_{\rm b}}{2 \pi} \frac{r_{\rm b}}{r(r+r_{\rm b})^{3}},
    \end{equation}
    
    \noindent where $r$ is the spherical radial coordinate.

    \item a gaseous CND with $M_{\rm CND}=10^8\msun$ and scale radius $R_{\rm CND}=50$~pc, defined by the surface density profile
    
    \begin{equation}
    \Sigma_{\rm CND}(R)=\frac{M_{\rm CND}}{2 \pi R_{\rm CND}^{2}} \exp \left(-\frac{R}{R_{\rm CND}}\right),
    \end{equation}
    
    \noindent where $R$ is the cylindrical radius.
    
    The vertical component and velocity field were computed as described in \cite{Lupi_et_al_2015}, using an iterative procedure that takes into account the global potential of the three components (stars, gas, and BH) and ensures vertical hydrostatic equilibrium.\footnote{For this purpose, the publicly available code {\textsc{gd\_basic}} (\url{http://www.dfm.uninsubria.it/alupi/software.html}) was used.} In our tests, the CND initially lays in the \textit{xy}-plane and, during the simulations, never significantly changes its orientation. The direction ($\theta$, $\phi$)=(0, 0) identifies the positive \textit{z}-axis, whereas ($\theta$, $\phi$)=($\pi/2$, 0) identifies the positive \textit{x}-axis.

    \item a central BH with initial mass $M_{\rm BH,0}=10^7\msun$.

\end{itemize}

Our tests were conducted at two different resolutions, as detailed in Table~\ref{tab:resolution}. We compared the high- and low-resolution results and verified that there is no significant difference in the system's evolution. Hence, in the following, we present only the high-resolution results.

The gas was initialized following an ideal-gas equation of state with $\gamma=5/3$ and a uniform temperature of $2\times 10^4$~K. Since, as described in \citet{Lupi_et_al_2015}, some spiral arms have to be dissipated and the disc undergoes an initial re-adjustment, we evolved our model in isolation with only gravity and hydrodynamics active for 20~Myr and reset the time to 0 at the end of this relaxation period. Afterwards, the tests with our model were performed with $\gamma=7/5$, to effectively mimic a mild cooling and favour gas accretion towards the centre, without actually using a cooling model (see e.g. \citealp{Dotti_et_al_2009}).

In all our simulations, we set $N_{\rm ngb} = 32 $ and $f_{\rm BH}=3$. We further assumed no boost factor in the Bondi accretion rate (i.e. $\alpha_{\rm acc} = 1$ in Equation~\ref{eq:bondi}) and fixed $r_{\rm accr,max} = 10$~pc, which, in our setup, is $\sim$$10\Delta x$, where $\Delta x$ is the mean inter-particle distance amongst gas particles.

The remaining parameters of our model vary depending on the test and are therefore quoted in the following sections.


\subsection{Comparing different feedback geometries}\label{sec:feedback_prescription_comparison}

Before testing our new accretion model (described in Section~\ref{sec:discaccr} and in \citealt{Cenci_et_al_2021}), we validated our new feedback model (presented in Section~\ref{sec:conical_feedback}), using the original accretion scheme by \citet{Angles-Alcazar_et_al_2017} and comparing their radial and collimated feedback geometries to our conical feedback prescription, for two different values of the kick velocity's magnitude: $v_{\rm out} = 10^3$ and $10^4$~km~s$^{-1}$ (see Table~\ref{tab:table2} for a summary of our tests). In this case, $\dot{M}_{\rm acc-BH}$ is computed using Equation~\eqref{eq:MdotaccBH_alcazar}, $\eta$ is fixed to 0.1 in the parameters file, and the feedback reference direction (for the collimated and conical cases) is computed on a time-step basis as parallel to the angular momentum of the gas within the BH kernel (hereafter $\bm{J}_{\rm gas}= J_{\rm gas}\cdot\bm{j}_{\rm gas}$, where $J_{\rm gas}$ and $\bm{j}_{\rm gas}$ are the magnitude and direction of the vector, respectively). Moreover, we assume a momentum-conserving scenario, by setting $p_{\rm b} = 1$. Hence, the values of $f$ as defined is Equation~\eqref{eq:f} are $f=0.032$ and $0.25$ for $v_{\rm out} = 10^3$ and $10^4$~km~s$^{-1}$, respectively. The cone semi-aperture $\hat\theta$ is set to $\pi/4$. 

\begin{figure*}
\centering
\includegraphics[width=1.0\textwidth]{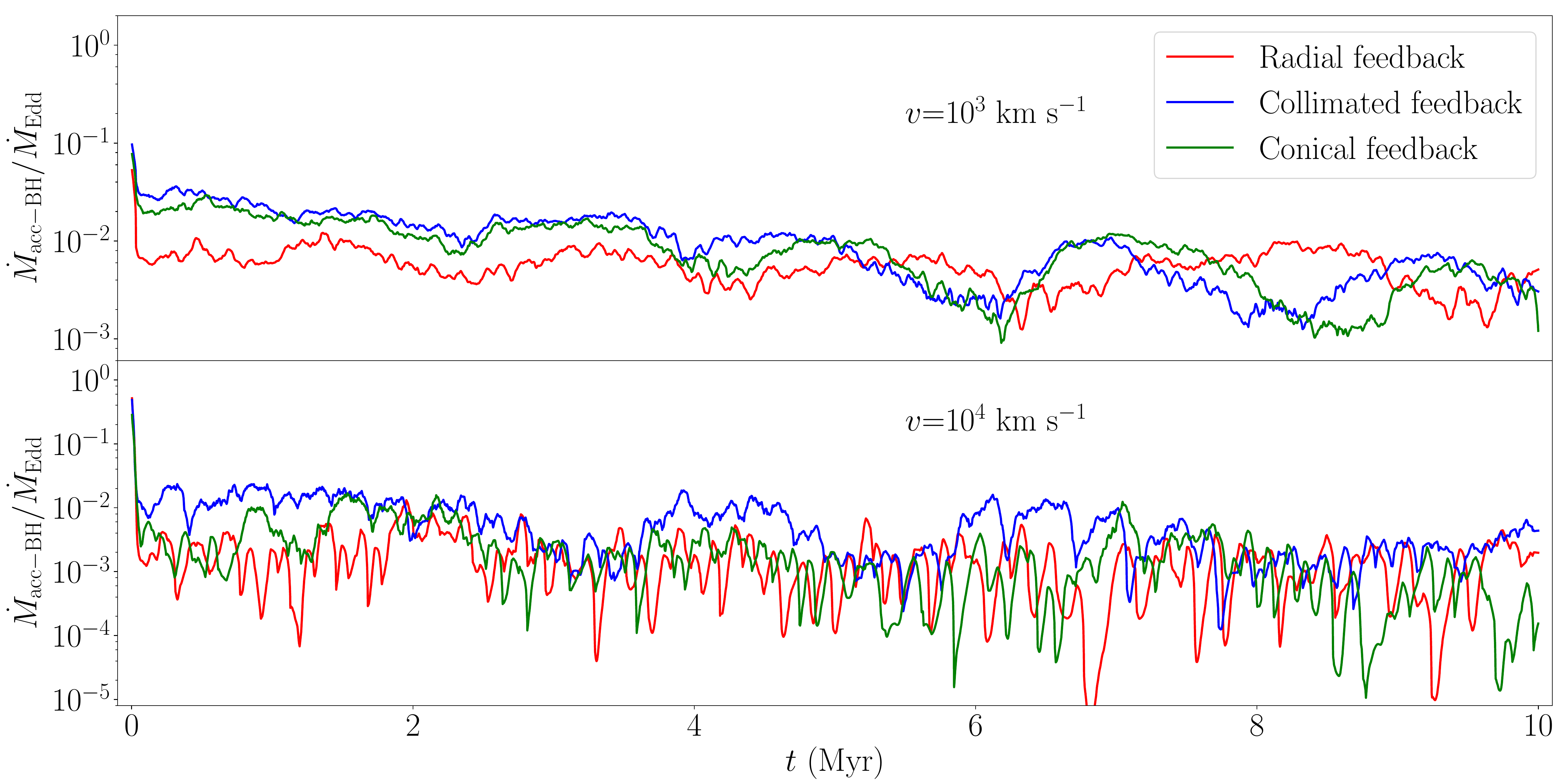}
\caption{Eddington ratio versus time (smoothed over 0.05~Myr) for three different feedback prescriptions -- radial (red line), collimated (blue), and conical (green) -- and for two different kick velocities -- $v_{\rm out}=10^3$ (top panel) and $10^4$~km~s$^{-1}$ (bottom panel), using the original accretion scheme by \citet{Angles-Alcazar_et_al_2017}. Note the different $y$-axis range in the two panels. In the case with $v_{\rm out}=10^3$~km~s$^{-1}$, in the first $\sim$4~Myr the Eddington ratio for the conical and collimated case is higher than in the radial case, where the larger interaction with the CND more effectively suppresses the accretion. In the following $\sim$6~Myr (and during the entire simulation with $v_{\rm out}=10^4$~km~s$^{-1}$), the main effect is that the accretion increases and decreases repeatedly due to the interaction between BH and overdensities that form during the simulations, also as a result of the stochastic kicks.}
\label{fig:eddratio_allmethods}
\end{figure*}

The Eddington ratio for these tests is shown in Figure \ref{fig:eddratio_allmethods}. Since the BH mass changes by less than one per cent over the whole 10 Myr of the simulations, the change in the Eddington ratio is approximately proportional to the BH accretion rate. At the very beginning of the simulations, accretion is quite high, because the BH is surrounded by a dense gas region. As soon as feedback becomes efficient, the accretion rate quickly drops. After this initial transient, which lasts $\sim$0.05~Myr, we have two different scenarios, depending on the kick velocity's magnitude. 

\begin{figure*}
\centering
\includegraphics[width=0.80\textwidth]{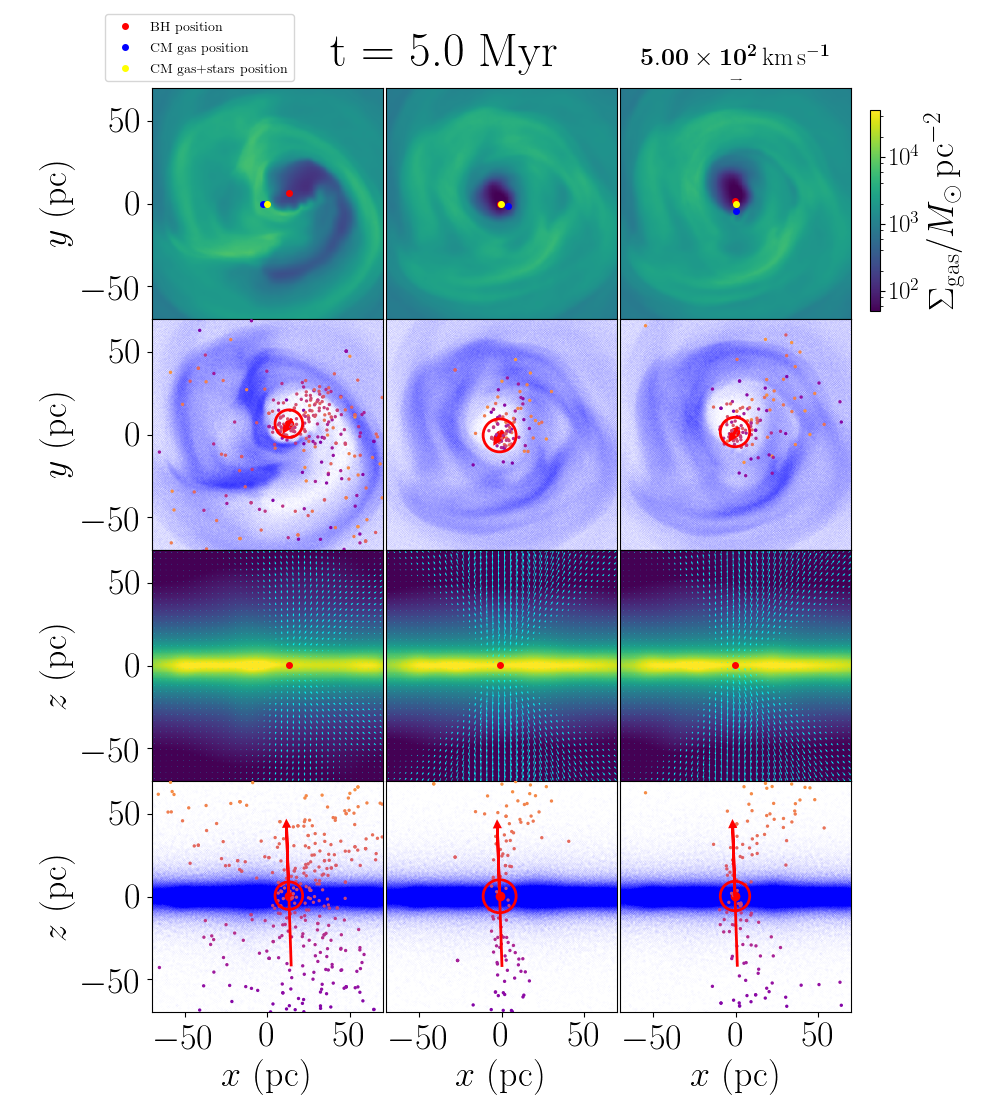}
\caption{Gas maps at 5~Myr for the tests with BH accretion modelled following \citet{Angles-Alcazar_et_al_2017}, and BH feedback with $v_{\rm out}=10^3\;\rm km\;s^{-1}$ and radial (left-hand column), collimated (central), and conical (right-hand) geometry. First row: gas surface density (viewed face-on); the red, blue, and yellow dots indicate the position of the BH, the centre of mass of the gas, and the centre of mass of gas+stars, respectively. Second row: gas particles' positions map (viewed face-on); the particles that received a kick are highlighted with the large dots, whose colour code reflects the particle's \textit{z} coordinate (greater values are lighter); the red circle shows the extent of the BH accretion length, whereas the red arrow shows the projection of the angular momentum direction of the gas in the BH kernel, arbitrarily normalized to be clearly visible. Third and fourth rows: same as the first and second rows, respectively, but with the CND seen edge-on (omitting the centre of mass positions) and with the gas velocity field overlaid to density map. In the radial case, the particles are heavily scattered when kicked along the disc plane, and a clear preferential direction of the particles after their launch is not recognizable. On the other hand, in the collimated and conical cases, after the particles have been launched, they define and maintain a conical shape. When the conical prescription is adopted, we can control the scatter in the resulting directions. The conical shape is still visible in the collimated case because, even if the particles are always launched in the perpendicular direction to the disc plane, without any direction distribution, the gas can freely expand in the low-density region above/below the disc, where there is not enough pressure to keep the wind confined.}
\label{fig:prescription_comparison_mosaic_close_up}
\end{figure*}

When $v_{\rm out} = 10^3$~km~s$^{-1}$, for the first $\sim$4~Myr the collimated feedback has the higher accretion rate, that in the conical case is slightly lower, and that in the radial case is much lower. This is due to the fact that the accretion increases as the interaction with the CND decreases: when particles are mostly pushed towards the CND (as in the radial model), accretion is impeded and therefore is lower than when particles are mostly pushed in the vertical direction (as in the collimated and conical cases), leaving the accretion flow from the CND almost unaltered. Furthermore, the accretion rate in the conical case is slightly lower than in the collimated case because as the semi-aperture increases there are more particles interacting with the CND. After around 4~Myr, the pattern becomes more complex. This is due to the formation of asymmetric gas overdensities (visible in the face-on density panels of Figure~\ref{fig:prescription_comparison_mosaic_close_up}, shown at 5~Myr) and the BH starts wandering as a result of the gravitational interactions with such overdensities. At later times, the BH can get close enough to these denser regions, such that the accretion rate and, consequently, feedback increase. This in turn pushes away the gas and decreases the accretion rate. This behaviour repeats, but it is heavily affected by where and when the overdensities form, which in turn is also influenced by the kicks that are intrinsically stochastic.

When $v_{\rm out} = 10^4$~km~s$^{-1}$, the fraction $f$ defined by Equation~\eqref{eq:f} is almost one order of magnitude higher than in the previous case (for a fixed $p_{\rm b}$ and $\eta$). Hence, to have the same accreted mass, fewer particles are needed. As a result, the $v_{\rm out}=10^4 \;\rm km\;s^{-1}$ case is more stochastically affected by fewer, higher-velocity particles that periodically lower the accretion by a significant amount, whereas in the other case ($v_{\rm out} = 10^3$~km~s$^{-1}$) the pattern is smoother. Each single particle is able to heavily modify the gas distribution around the BH in a specific direction in the CND plane. All the three tests with different kick's velocity magnitude exhibit the behaviour in which the accretion rate increases and decreases repeatedly due to the interaction between BH and overdensities.

We note that the accretion rate in the $v_{\rm out} = 10^4$~km~s$^{-1}$ case is typically lower than that in the $v_{\rm out} = 10^3$~km~s$^{-1}$ case. It is however hard to generalize this result, given the short time-scale and the extremely idealized setup of the simulation, which result in a negligible BH mass growth. We plan to investigate such trend in future works.

Figure~\ref{fig:eddratio_allmethods} shows that $\dot{M}_{\rm acc-BH}/\dot{M}_{\rm Edd}\lesssim0.01$ even if we are using a feedback model that aims to represent a fast wind in a radiatively efficient accretion phase. However, the accretion rate depends mostly on the properties of the CND, which have been chosen to be in quasi-equilibrium, in order to prevent disc fragmentation, the need for additional physics such as star formation and its associated feedback. This choice allowed us to isolate the effect of the AGN feedback prescription, at the cost of having low accretion rates.

We verified that the kicked particles, for the collimated and conical cases, define a conical shape far from the launch region for the entire simulation, as illustrated in the bottom panels of Figure~\ref{fig:prescription_comparison_mosaic_close_up} for the $v_{\rm out} = 10^3$~km~s$^{-1}$ case, where the particles that have been given a kick are highlighted with the large dots. The semi-aperture parameter determines the large-scale particle direction dispersion, because it reflects the small-scale launch directions distribution. However, as soon as the wind leaves the high-density region of the CND, where gas pressure helps keeping it collimated, the gas expands in the radial direction, forming a cone (also in the collimated case).

\begin{figure*}
\centering
\includegraphics[width=1\textwidth, trim=0 0 0 0, clip]{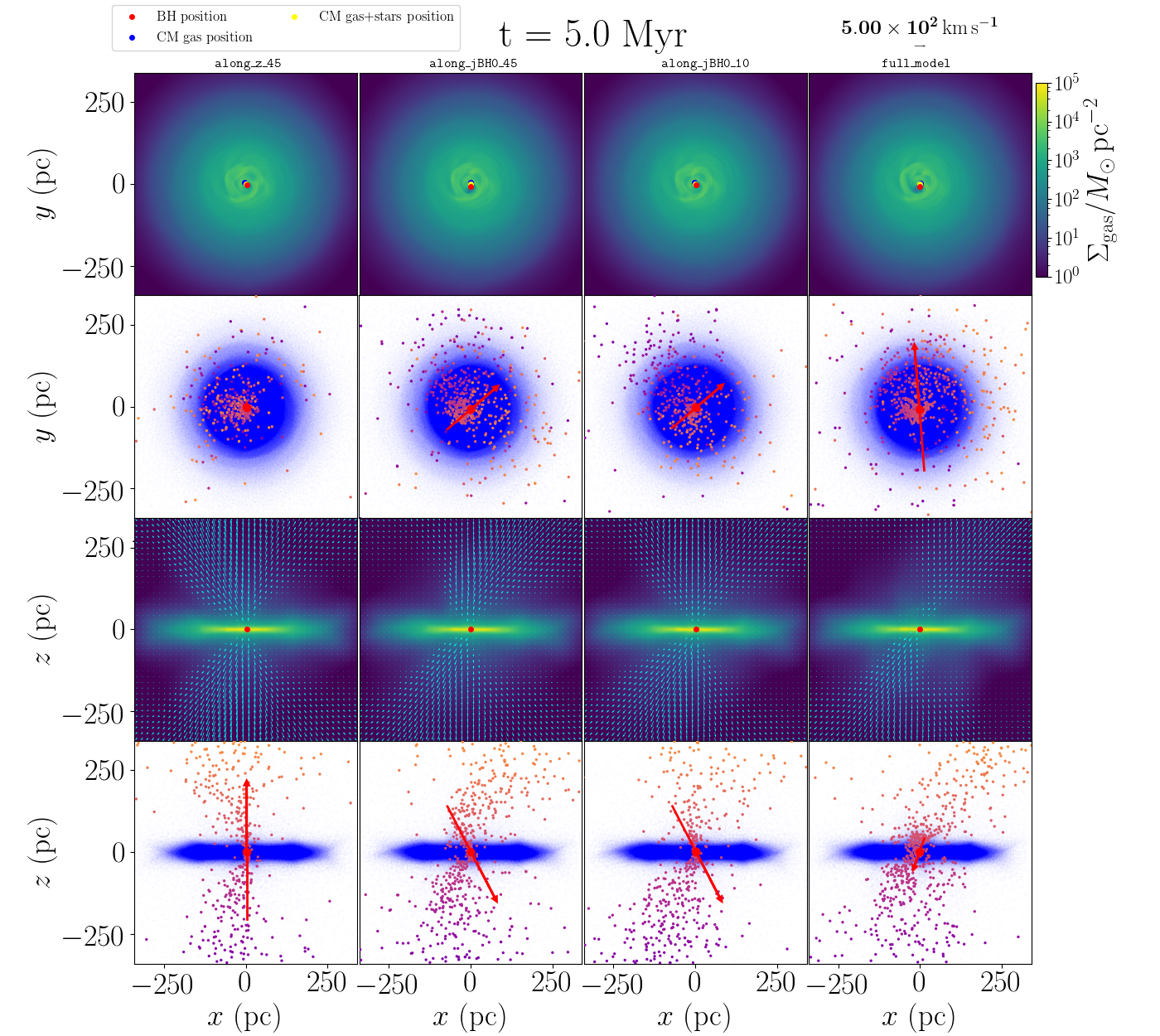}
\caption{Same as Figure~\ref{fig:prescription_comparison_mosaic_close_up}, but for the tests described in Sections~\ref{sec:control_models} and \ref{sec:complete_model}. From left to right: \texttt{along\_z\_45}, \texttt{along\_jBH0\_45}, \texttt{along\_jBH0\_10}, \texttt{full\_model}. Considering \texttt{along\_z\_45}, the kicked particles at large scales (large dots in the bottom panels) define a conical shape, which is maintained for the whole duration of the tests. Tests \texttt{along\_jBH0\_45} and \texttt{along\_jBH0\_10} show no significant dependence of the result on the semi-aperture. Regarding \texttt{full\_model}, the red arrow highlights that the BH spin, at $\sim$5~Myr, is directed along the disc plane, hence the large-scale outflow shape is more affected by the interaction of the wind with the disc (the particle distribution looks more cluttered, especially near the centre).}
\label{fig:mosaic-controls-bench}
\end{figure*}

\begin{figure*}
\centering
\includegraphics[width=1\textwidth, trim=0 0 0 0, clip]{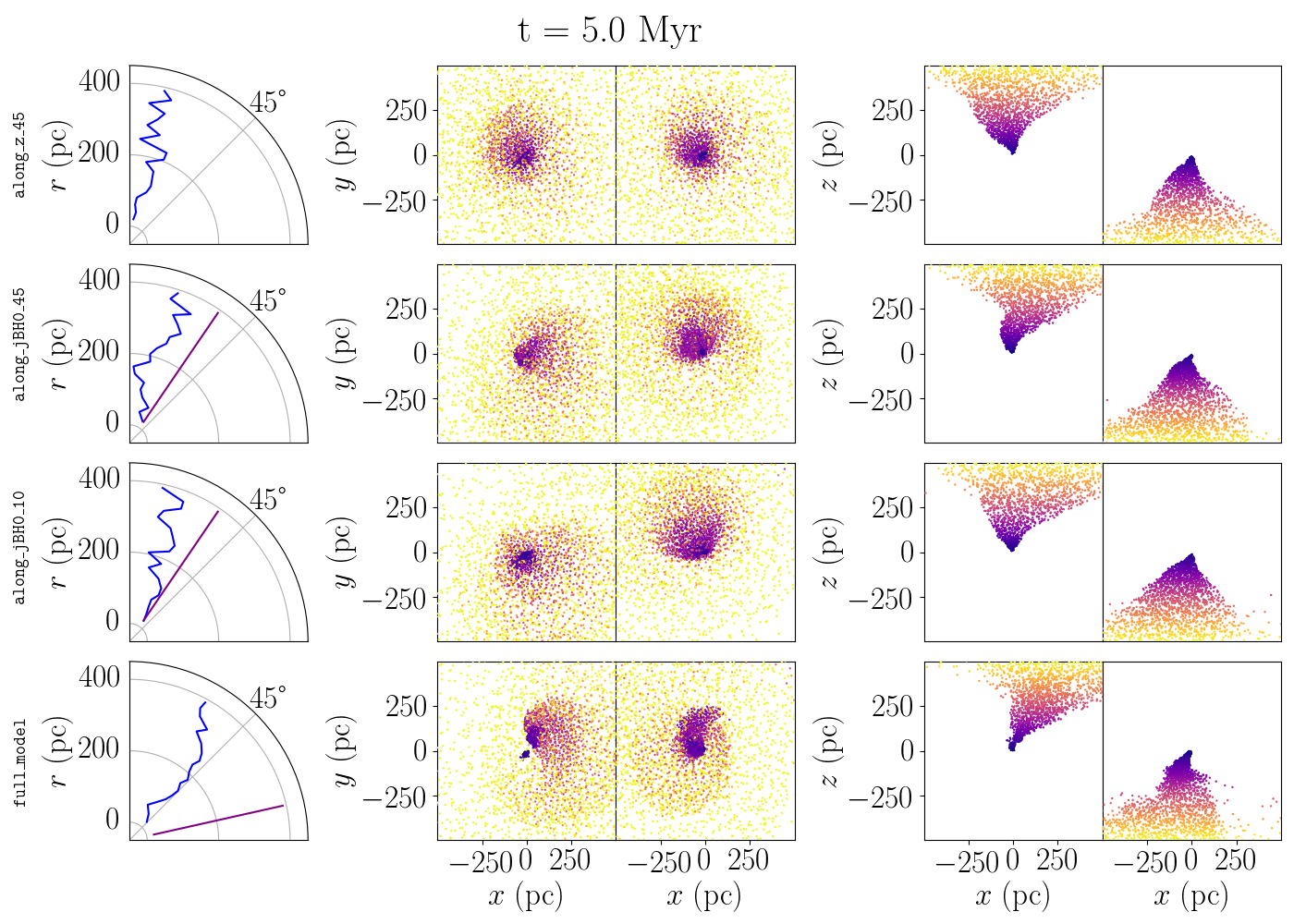}
\caption{From top to bottom, respectively: \texttt{along\_z\_45}, \texttt{along\_jBH0\_45}, \texttt{along\_jBH0\_10}, and \texttt{full\_model} at 5~Myr. Left-hand panels (polar plots): $\bar\theta$ (blue line) and instantaneous feedback direction (purple line) as a function of radius, plotted for the snapshot. Central panels: $xy$-projections of the upwards (left) and downwards (right) outflow particles positions. Right-hand panels: $xz$-projections of the upwards (left) and downwards (right) outflow particles positions. The colour code reflects the particle \textit{z} coordinate (lighter are further from the $xy$-plane). The polar plots show that the direction of the resulting outflow $\bar\theta$ varies with the distance from the BH and is not exactly equal to its launch value (shown as the purple line). The visual appearance of the outflow (cartesian panels) is also quite similar.}
\label{fig:mosaic-controls-bench-polar}
\end{figure*}

\begin{figure*}
\centering
\includegraphics[width=1\textwidth, trim=0 0 0 0, clip]{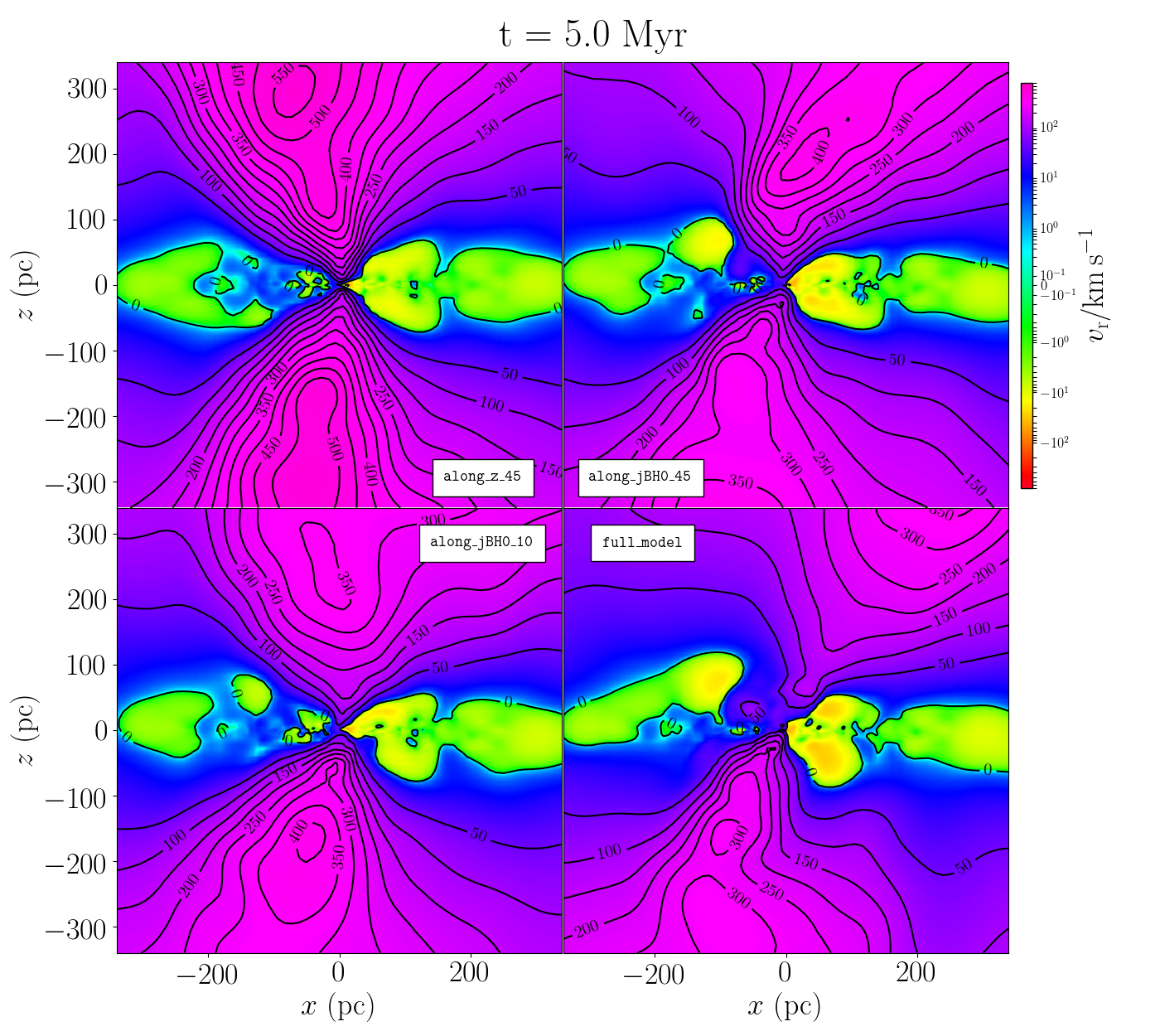}
\caption{Contour plots of the radial velocity of the gas component in the simulations, from a snapshot at 5.0~Myr and for the four tests described in Sections \ref{sec:control_models} and \ref{sec:complete_model}. An inflow from the disc plane is quite evident, but note that its velocity is two orders of magnitudes lower than that of the outflow. The kicked gas particles shock with the gas in the disc immediately after being launched. The net effect is that (i) they slow down, and (ii) they are scattered off their initial trajectory. These effects are more prominent when the particles are launched in a direction closer to the disc plane. Hence, from the case with the outflow along the \textit{z}-axis, to the case where the direction is slightly tilted (\texttt{along\_jBH0\_45} and \texttt{along\_jBH0\_10}), to the case where the BH spin crosses the disc plane, the maximum outflow speed decreases.}
\label{fig:contours-plot}
\end{figure*}

\begin{figure*}
\centering
\includegraphics[width=1\textwidth, trim=0 0 0 0, clip]{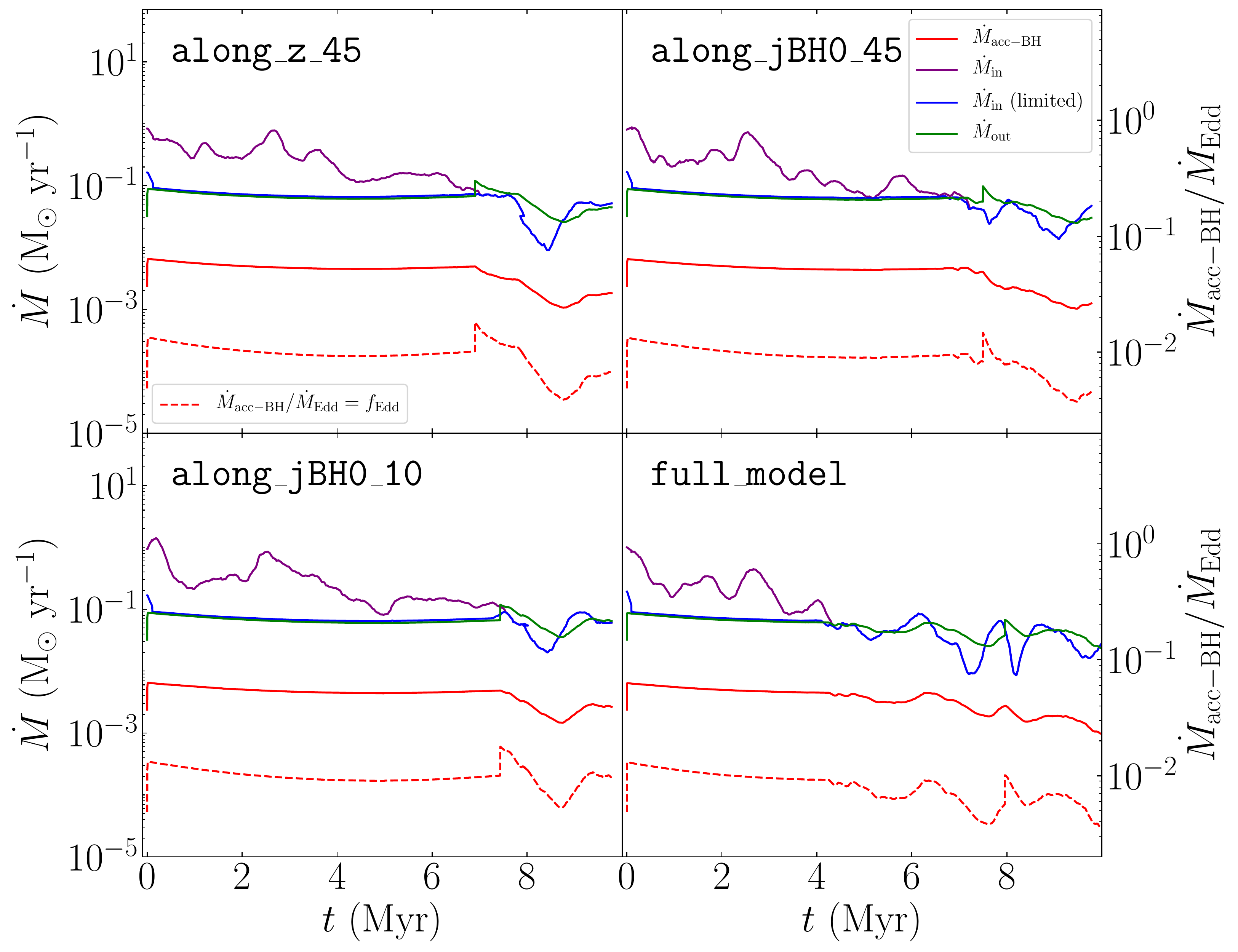}
\caption{Mass rates (solid lines) and BH Eddington ratios (dashed lines) for the evolving components of our accretion model, for the four tests described in Sections \ref{sec:control_models} and \ref{sec:complete_model} ($\dot{M}_{\rm acc-BH}$ computed as in Equation~\ref{eq:fedd}, $\dot{M}_{\rm in}$ as in Equation~\ref{eq:bondi}, $\dot{M}_{\rm in}$ after the limit at the self-gravitating mass has been applied, and $\dot{M}_{\rm out}$ as in Equation~\ref{eq:mom_load}). The inflow rate from large scales of run \texttt{full\_model} starts to decrease at earlier times, compared to the other tests. This occurs when the BH spin direction crosses the disc plane. Moreover, the sudden increase in $\dot{M}_{\rm out}$ (at $\sim$7~Myr) is caused by the sudden change in the value of the radiative efficiency due to the transition between prograde and retrograde accretion in the sub-grid model (which evolves even when the direction is fixed, as explained in Section~\ref{sec:feedback_prescription_comparison}).}
\label{fig:rates-control-bench} 
\end{figure*}


\subsection{Testing the conical feedback model}\label{sec:control_models}

In this and the following section, we test both our new accretion model and conical feedback prescription.  For all these tests, the initial quantities of the sub-grid accretion disc model are $\eta_{0}=0.1$, $M_{\rm disc,0}=5\times 10^4 \msun$, $a_{\rm BH,0} = 0.5$, and

\begin{equation}
\bm{j}_{\rm BH,0}=
\left(
\begin{array}{c}
        \sin(\upi/6)\cos(\upi/4) \\
        \sin(\upi/6)\sin(\upi/4) \\
        -\cos(\upi/6)
\end{array}
\right).
\label{eq:jBH0}
\end{equation}

Two more quantities must be provided to the accretion disc model to be initialized properly: we set the initial Eddington ratio\footnote{The initial Eddington ratio $f_{\rm Edd, 0}$ is only needed to initialize the disc properties: during the first time-steps, $f_{\rm Edd}$ self-adjusts according to $\dot{M}_{\rm in}$.} $f_{\rm Edd, 0} = 5\times10^{-3}$ and the ratio of the circularization to self-gravitating radius $R_{\rm circ}/R_{\rm sg} = 0.5$ (for more details, see \citealt{Cenci_et_al_2021}).

Hence, the coordinates of the initial BH spin direction in our spherical coordinates system are equal to $(\theta_{0},\phi_{0})=(5\upi/6,\upi/4)$. This is such that the initial misalignment angle between the BH spin and the angular momentum of the resolved gas reservoir $\theta_{{\rm BH}-\rm gas,0} = \arccos\left(\bm{j}_{\rm BH,0}\cdot\bm{j}_{\rm gas,0}\right)$ is equal to $5\pi/6$, since $\bm{j}_{\rm gas,0}$ is aligned with the \textit{z}-axis.

For simplicity, we start by exploring how the choice of the cone's axis direction and semi-aperture affects the resulting outflow in a series of tests in which $\eta$ and $\dot{M}_{\rm acc-BH}$ are taken as an input from the accretion disc module, but the cone axis direction is fixed in time (although the BH spin evolves). We are particularly interested in assessing how different relative orientations between the feedback reference direction and the CND symmetry axis affect the outflow properties.

We tested the following three cases (see Table~\ref{tab:table2}):

\begin{enumerate}

    \item Cone's semi-aperture $\hat\theta=\upi/4$ and cone's axis direction parallel to the \textit{z}-axis (i.e. $\theta = \phi = 0$).

    \item $\hat\theta=\upi/4$ and cone's axis direction constant throughout the test and equal to the value of $\bm{j}_{\rm BH0}$ given by Equation~\eqref{eq:jBH0}.

    \item $\hat\theta=\upi/18$ and cone's axis direction as in the previous case.
    
\end{enumerate}

We forced the direction to be a definite value during these simulations, which is a reference-frame-dependent choice, because the CND starts in the \textit{xy}-plane and we verified that its orientation does not change significantly during the simulation under the effect of feedback. In all three cases, as well as in the ``full model'' test described in the following section, we assume a momentum-conserving scenario (i.e. $p_{\rm b} = 1$) and a kick velocity magnitude $v_{\rm out} = 10^3$~km~s$^{-1}$.

The test \texttt{along\_z\_45}, in which feedback is exerted along the $z$-axis, is similar in nature to the conical-feedback case of the previous section, except for the mass rates, which are now computed using our sub-grid accretion disc model. The first column of Figure \ref{fig:mosaic-controls-bench} looks very similar in every aspect to the last column of Figure~\ref{fig:prescription_comparison_mosaic_close_up}, as expected, since the accretion model (the only difference between the two cases) affects the mass rates (hence the number of selected particles) but it has no relation with the outflow geometrical features. Moreover, the kicked particles at large scales (large dots in the bottom panels) define a conical shape, which is maintained for the whole duration of the tests.

We further defined the outflow as the subset of gas particles which have $v_{\rm radial}>2\times\langle v_{\rm tangential}\rangle$ (where $\langle v_{\rm tangential}\rangle$ is the mean tangential velocity over the CND and maintains an approximately constant value of $\sim$85~km~s$^{-1}$), and that are at a radius greater than the BH accretion length. We divided the outflow in spherical shells centred on the BH position and, for the particles belonging to each shell, we computed the median position, which corresponds to a direction $(\bar\theta)$ in spherical polar coordinates. This direction, computed shell by shell, is therefore dependent on the distance from the BH.

Considering the blue line in the polar plot of the top row of Figure~\ref{fig:mosaic-controls-bench-polar}, the outflow direction $\bar\theta$ deviates at every radius from being exactly on the \textit{z}-axis, because of two concurrent effects: the particle scattering effects when interacting with the surrounding disc and the pressure gradients that develops due to hydrodynamics, since the wind is expanding in the low-density region above/below the disc. 

Comparing tests \texttt{along\_jBH0\_45} and \texttt{along\_jBH0\_10} (see the two central columns of Figure~\ref{fig:mosaic-controls-bench} and the two central rows of Figure~\ref{fig:mosaic-controls-bench-polar}), which only differ with each other in the value of the semi-aperture $\hat{\theta}$, we note that this difference has no significant effect on the resulting outflow. As in the case of test \texttt{along\_z\_45}, the direction of the resulting outflow $\bar\theta$ varies with the distance from the BH and is not exactly equal to its launch value (shown as the purple line in Figure~\ref{fig:mosaic-controls-bench-polar}).

In Figure~\ref{fig:contours-plot}, we show the velocities' magnitude of the outflowing gas. In test \texttt{along\_z\_45} (top-left panel), the speed reaches more than $\sim$550~km~s$^{-1}$, whereas in tests \texttt{along\_jBH0\_45} and \texttt{along\_jBH0\_10} (respectively, top-right and bottom-left), it reaches $\sim$400~km~s$^{-1}$. This effect can be explained considering that the kicked gas particles shock with the gas in the disc immediately after being launched. The net effect is that (i) they slow down, and (ii) they are scattered off their initial trajectory. This effect is more prominent as the particles are launched in a direction closer to the disc plane. 

The minor differences in the structure and velocity of the outflows of the three cases presented in this section are consistent with the very small discrepancies observed in the mass accretion rates computed in the three runs, as shown in Figure~\ref{fig:rates-control-bench}. Overall, some qualitative differences can be seen in the kicked particles distribution up to 300 pc, as illustrated in the bottom panels of Figure~\ref{fig:mosaic-controls-bench}. However, when we select the particles according to the radial velocity criterion explained above, those differences become far less evident and the three tests appear similar in their outflow direction $\bar\theta$ and their qualitative appearance, up to 400 pc (see Figure~\ref{fig:mosaic-controls-bench-polar}). The main difference remains instead the outflow maximum speed, that is greater when the particles interact less with the disc. 


\subsection{The full model}\label{sec:complete_model}

In this section, we present the benchmark run, in which we coupled our sub-grid feedback model with the self-consistent spin evolution model presented in \cite{Cenci_et_al_2021}, as done already for the tests of Section~\ref{sec:control_models}, additionally modelling an outflow cone axis parallel to the BH spin direction vector, which is evolving with time (as opposed to the cases of the previous section). The initial accretion disc values for this simulation are the same listed in the previous section. The results are shown in Figures~\ref{fig:mosaic-controls-bench}--\ref{fig:rates-control-bench}, for an easier comparison with the previous tests.

At around 4~Myr, the BH spin direction starts to be along the CND plane. Since most of the particles are kicked in that direction, the gas is kept far from the BH and, looking at the bottom-right panel of Figure~\ref{fig:rates-control-bench}, the inflow rate from large scales starts to become lower at earlier times, compared to the other tests. Note that the prescription for the accretion disc evolution limits the disc growth when the inflow rate satisfies our criterion, hence the blue line starts to overlap with the purple one only when the inflow rate is low enough and the disc mass has decreased below the self-gravitating mass. 

Furthermore, the large-scale outflow shape is heavily affected by the interaction with the CND, as visible in the bottom-right panel of Figure~\ref{fig:mosaic-controls-bench}. The polar plot in the last row of Figure~\ref{fig:mosaic-controls-bench-polar} (blue line) shows that the outflow is slightly closer to $\upi/4$ than in the other three runs. None the less, the outflow direction near the origin, which should be aligned with the launch direction, is already far from it and closer to the \textit{z}-axis, due to the strong interaction with the disc. Looking at the bottom-right panel of Figure~\ref{fig:contours-plot}, the maximum velocity's magnitude of the outflowing material in this case lowers to $\sim$300~km~s$^{-1}$. In this case indeed, the particles, following the BH spin direction, start to be pushed against the disc plane after $\sim$4~Myr, slowing down even more than in the previous cases. 


\section{Discussion and conclusions}\label{sec:conclusions}

We designed a novel BH accretion and momentum-driven feedback model and interfaced it with the publicly available code {\textsc{gizmo}}. Our prescription is based on that described in \cite{Angles-Alcazar_et_al_2017}, which we significantly modified. We take into account the presence of a sub-grid accretion disc and ensure self-consistency between its mass evolution, the inflow mass rate from large resolved scales, the outflow mass rate, and the accretion rate on to the central BH. A set of particles is stochastically selected within the BH kernel length, and a fraction of the particles' mass is added to the BH mass. These particles are then kicked (with a velocity magnitude defined by the user) in such a way that the direction is randomly extracted in a cone of a given aperture, whose axis can be chosen arbitrarily, according to the underlying physical assumptions of the model employed. Finally, we coupled the conical feedback model with the accretion disc prescription described in \cite{Cenci_et_al_2021}, wherein the accretion rate on to the BH depends on the inflow rate but is mediated by the accretion disc. Furthermore, we set the reference direction for the conical feedback parallel to the BH spin direction. 

There are some caveats in our model that we will address in the future, in the aim at further improving the model. Specifically, \textit{(i)} when the mass of the sub-grid disc+BH system decreases (due to the decrease in mass of the accretion disc), the BH dynamical mass remains constant, \textit{(ii)} while statistical properties would be the same, carrying out a new set of simulations and changing the random number generator seeds would produce different positions in the overdensities when the number of particles which exert feedback is small. In addition, particularly in misaligned conditions when the outflow impinges into the CND, the asymmetries could become artificially more pronounced for lower mass resolution in the outflow. Using only the particles which are already present in the simulation to exert feedback restricts the possibilities of implementation to mitigate the above-mentioned effect.
A solution could be to generate ``wind particles'' very close to the BH (to minimise discreteness), which are responsible to create the feedback effect and to which redistribute the dynamical mass in excess \citep[see e.g.][]{Cuadra_et_al_2006,Pillepich_et_al_2018,Torrey_et_al_2020}

We carried out a suite of hydrodynamic tests on a CND, comparing the conical feedback prescription with the collimated and radial feedback geometries described in \cite{Angles-Alcazar_et_al_2017}. We started by exploring how the choice of the cone's axis direction and semi-aperture affects the resulting outflow. Then we performed a test with our complete implementation, using an outflow cone axis parallel to the BH spin direction vector, which evolved with time. 

We summarize the findings of our \textit{N}-body, hydrodynamic tests as follows. Comparing different feedback geometries, we verified that all our tests exhibit the formation of gas overdensities, which interact with the BH and make the accretion behaviour complex (see Figure~\ref{fig:prescription_comparison_mosaic_close_up}). In the radial case, the particles are heavily scattered when kicked in the direction of the CND plane, and the exerted feedback lowers the accretion considerably with respect to the collimated and conical cases (see Figure~\ref{fig:eddratio_allmethods}). In the collimated and conical cases, the kicked particles define and maintain a conical shape.

Analysing the influence of the cone's axis direction and semi-aperture, as well as the full coupling with the BH spin evolution model, we find that:

\begin{itemize}

\item The shape and orientation of the outflows are similar, regardless of the different orientations and semi-apertures considered in this work (see Figures~\ref{fig:mosaic-controls-bench} and \ref{fig:mosaic-controls-bench-polar}). As a consequence, also the mass inflows and outflows are similar (see Figure~\ref{fig:rates-control-bench}).
    
\item A conical shape can always be identified, even at late times, but the outflow's structure is complex (see Figures \ref{fig:mosaic-controls-bench}--\ref{fig:contours-plot}). In particular, the outflowing material in the central parts of the outflow has a high speed, which decreases with the distance from the cone's axis (see Figure~\ref{fig:contours-plot}). Moreover, it is not straightforward to relate the launching direction and the resulting outflow orientation (the latter depending also on the distance from the BH), because the interaction with the CND plays a dominant role in shaping the outflow (see Figure~\ref{fig:mosaic-controls-bench-polar}).
    
\item The more the cone's axis is tilted in the direction of the CND plane, the more the kicked particles are deviated and slowed down (see Figure~\ref{fig:contours-plot}), and the large-scale inflow is decreased (see Figure~\ref{fig:rates-control-bench}). This is particularly evident when the outflow crosses the CND.
    
\end{itemize}

In conclusion, our model is able to self-consistently capture the basic physics of accretion discs and the wind driving mechanism, also accounting for the spin-disc coupling and its effect onto the outflows. It is flexible enough to allow the user to choose a preferential direction for the wind, if desired. This model can be adopted as a sub-grid prescription to multiscale simulations and in future works we plan to employ it to investigate how the spin-feedback coupling affects both the BH growth in mass, the spin evolution, and the interplay between the BH and the ISM of the galaxy host.

\section*{Acknowledgements}
We acknowledge the CINECA award under the ISCRA initiative, for the availability of high-performance computing resources and support (projects numbers HP10CFXS9S and HP10CRRSO8). LS acknowledges support from `BiD4BEST' - European Innovative Training Network (ITN) funded by the Marie Sk\l{}odowska-Curie Actions (860744) in Horizon 2020. AL acknowledges support from the European Research Council Advanced Grant N. 740120 `INTERSTELLAR'. This work reflects only the authors' view and  the  European Research Commission is not responsible for information it contains. MD acknowledges funding from MIUR under the grant PRIN 2017-MB8AEZ. The analyses reported in this work have been mainly performed using \textsc{pynbody} \citep{pynbody}.

\section*{Data Availability Statement}
The data underlying this article will be shared on reasonable request to the corresponding author.

\scalefont{0.94}
\setlength{\bibhang}{1.6em}
\setlength\labelwidth{0.0em}
\bibliographystyle{mnras}
\bibliography{non-isotropic_feedback_from_accreting_spinning_black_holes}
\normalsize

\bsp 
\label{lastpage}
\end{document}